\documentclass[conference]{IEEEtran}
\IEEEoverridecommandlockouts

\usepackage{cite}
\usepackage{amsmath,amssymb,amsfonts}
\usepackage{algorithmic}
\usepackage{algorithm}
\usepackage{graphicx}
\usepackage{textcomp}
\usepackage{xcolor}
\usepackage{subfig}
\usepackage{bm} 
\usepackage{balance}
\usepackage{epstopdf}

\def\BibTeX{{\rm B\kern-.05em{\sc i\kern-.025em b}\kern-.08em
    T\kern-.1667em\lower.7ex\hbox{E}\kern-.125emX}}

\begin{document}

\title{Resilient and Freshness-Aware Scheduling for Industrial Multi-Hop IAB Networks: A Packet Duplication Approach\\
\thanks{This work was supported in part by the Fundamental Research Funds for the Central Universities (Collaborative Inno-vation Center of Railway Trafﬁc Safety) under Grant 2025JBXT010; in part by the National Natural Science Foundation of China under Grant U2368201, and Grant 62221001; in part by the Project of China Railway Corporation under Grant K2025X007.}}

\author{
    \IEEEauthorblockN{Shuo Zhu\IEEEauthorrefmark{1}, Siyu Lin\IEEEauthorrefmark{1}, Zijing Wang\IEEEauthorrefmark{1}, Qiao Ren\IEEEauthorrefmark{1}, Xiaoheng Deng\IEEEauthorrefmark{2}, and Bo Ai\IEEEauthorrefmark{1}}
    \IEEEauthorblockA{\IEEEauthorrefmark{1}\textit{School of Electronic and Information Engineering}, \textit{Beijing Jiaotong University}, Beijing 100044, China \\
    \IEEEauthorrefmark{2}\textit{School of Electronic Information}, \textit{Central South University}, Changsha 410083, China \\}    
    \IEEEauthorblockA{Siyu Lin (Corresponding Author) Email: sylin@bjtu.edu.cn}
}

\maketitle

\begin{abstract}
In industrial millimeter-wave (mmWave) multi-hop Integrated Access and Backhaul (IAB) networks, dynamic blockages caused by moving obstacles pose a severe threat to robust and continuous networks. While Packet Duplication (PD) enhances reliability by path diversity, it inevitably doubles the traffic load, leading to severe congestion and degraded Age of Information (AoI). 
To navigate this reliability-congestion trade-off,  we formulated an optimization problem in a multi-hop IAB scenario that minimizes the average AOI while satisfying strict queue stability constraints. We utilize Lyapunov optimization to transform the long-term stochastic optimization problem into tractable deterministic sub-problems. To solve these sub-problems efficiently, we propose a Resilient and Freshness-Aware Scheduling (RFAS) algorithm. 
Simulation results show that in blockage-prone environments, RFAS significantly outperforms baselines by maintaining a Packet Delivery Ratio (PDR) above 95\%. Crucially, it strictly guarantees queue stability under hard buffer constraints, whereas baselines suffer from buffer overflows. Furthermore, RFAS reduces the network load imbalance by 19\% compared to the baseline in high-frequency traffic scenarios. This confirms RFAS as a robust and sustainable solution for real-time industrial control loops.
\end{abstract}

\begin{IEEEkeywords}
Multi-hop Integrated Access and Backhaul (IAB) networks, Age of Information (AoI), Packet Duplication (PD), Lyapunov optimization, link scheduling.
\end{IEEEkeywords}


\section{Introduction}

\IEEEPARstart{T}{he} Industrial Internet of Things (IIoT) will enable a more resilient and adaptive manufacturing
paradigm in the Industry 5.0 era \cite{Mah2025Resilience}. 
In this new industrial landscape, the amount of data transmission between manufacturing machines and industrial information systems will rise dramatically, and the transmission of data directly affects the efficiency of industrial production. 
Millimeter-wave (mmWave) multi-hop wireless network based on Integrated Access and Backhaul (IAB) provides flexible coverage and high bandwidth for such environments. 
However, dynamic obstacles in the factory scenario can cause frequent blockages in directional mmWave links \cite{Sopin2024mmWave}, resulting in intermittent connections and outdated status updates. Consequently, a robust and continuous network architecture is required to ensure the stability of mission-critical operations.

To enhance transmission reliability, the 3rd Generation Partnership Project (3GPP) introduced the Packet Duplication (PD) strategy, which transmits critical data packets through disjoint paths \cite{Aij2019PD}. Although PD significantly enhances the resilience against single-path failures, it fundamentally violates the goal of resource efficiency and inevitably exacerbates congestion and queuing delays \cite{Ge2025PD,Ning2025CJE}. Therefore, there exists a tricky trade-off between the traffic load in the network and the freshness of the data.

Traditional scheduling problems mainly focus on throughput maximization or delay minimization, treating packets independently with constant value over time. However, in critical industrial control loops, the value of outdated data packets are less than fresh ones. To quantify this temporal value, Age of Information (AoI) has been widely established as a canonical metric for information freshness. 

In single-hop networks, the AoI-oriented problems has been well studied. In recent years, the research focus of researchers has shifted to multi-hop networks, especially in IAB networks. Existing literature has extensively explored relay-aided AoI minimization strategies \cite{Zhang2024Freshness, Wang2025RelayAOI,chen2026AOICJE}, multi-path scheduling designed for high-capacity transmission \cite{Wang2022MC,D2D2025CJE}, and energy-efficient resource allocation in IIoT \cite{Sho2023Sca,Feng2024QLearning}. 
However, the specific intersection of PD strategy and freshness-aware scheduling in IAB networks remains largely unexplored. Unlike the aforementioned studies that consider throughput capacity or energy savings, deploying PD for reliability introduces unique challenges. The redundant traffic imposes severe pressure on resource constrained multi-hop networks. Therefore, a critical gap remains in addressing the specific trade-off between reliability guarantees and the reduction in freshness caused by congestion. 

Addressing this gap, this paper investigates the balance between PD-induced congestion and information freshness in multi-hop IAB networks. We formulate a reliability-constrained AoI minimization problem and propose a low-complexity, Lyapunov-based Resilient and Freshness-Aware Scheduling (RFAS) algorithm. This algorithm effectively reconciles congestion with reliability, guaranteeing queue stability under hard constraints while maximizing network-wide freshness.

\section{Modeling and Problem Formulation}

\begin{figure}[t]
    \centering
    \includegraphics[width=0.9\columnwidth]{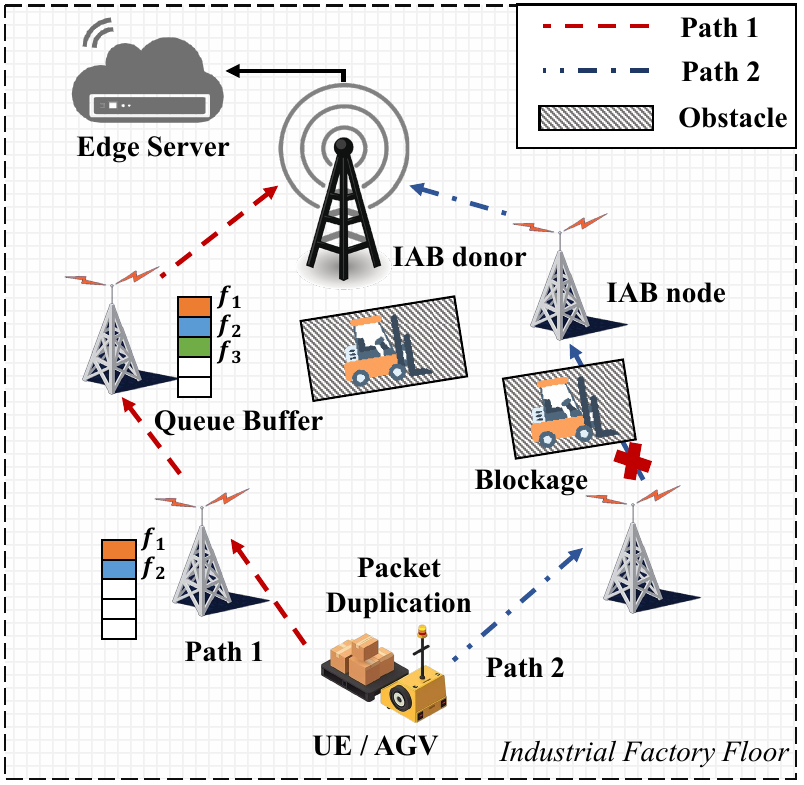} 
    \caption{System model of the resilient industrial IAB network. The UE utilizes PD over two disjoint paths to transmit updates amidst dynamic blockages, \textit{e.g.}, moving forklifts.}
    \label{fig:sys_model}
\end{figure}

This section presents the mathematical modeling and formulates the transmission scheduling problem for an industrial IAB network based on the PD strategy.

\subsection{Mathematical Modeling}
A typical scenario of a resilient industrial IAB network is shown in Fig.\ref{fig:sys_model}, which contains an IAB donor, $\mathit{V}$ IAB nodes, and $\mathit{U}$ User Equipments (UEs), \textit{e.g.}, AGVs, uniformly deployed within an industrial factory. The network topology is modeled as a directed graph $\mathcal{G} = \{\mathcal{N}, \mathcal{L}\}$, where $\mathcal{N} = \mathcal{V} \cup \mathcal{U}$, $\mathcal{V} = \{0, . . . , \mathit{V}\}$ represents the set of IAB nodes with the IAB donor indexed as 0, $\mathcal{U} = \{1, . . . , \mathit{U}\}$ represents the UE set, and $\mathcal{L} = \{l_{i,j} | i,j \in \mathcal{N}\}$ stands for the set of directed wireless links, where ${l_{i,j}}$ and ${l_{j,i}}$ denoting distinct unidirectional links.

Let $\mathcal{F}$ denote the set of critical industrial traffic flows. Each flow $f \in \mathcal{F}$ originates from a UE and eventually reaches the IAB donor. The source node and destination node of each flow be $s_f$ and $d_f$, respectively.

To satisfy the stringent reliability requirements, we adopt a disjoint PD strategy. For each flow $f$, two routing paths $P_{f} = \{p_{f,1}, p_{f,2}\}$ are pre-configured between $s_f$ and $d_f$. These paths do not overlap with each other except for the $s_f$ and the $d_f$, to ensure sufficient spatial isolation, so that the link interruption caused by dynamic blocking is difficult to simultaneously interrupt two wireless links. The data packets generated at the source node will be copied and simultaneously injected into both paths. If at least one copy of the data packet reaches the destination, it is considered to have been successfully transmitted. If the copied data packet arrives later, it will be discarded at the Packet Data Convergence Protocol (PDCP) layer. 
At the link layer, time slot based scheduling is employed. Let $T$ denote the total number of time slots. Each packet is time-stamped upon its generation. We assume that every packet can be transmitted within a single time slot.


\subsubsection{AoI modeling}
Let $A_{j,f}(t)$ denote the instantaneous AoI of flow $f$ at node $j \in \mathcal{N}$ at the beginning of slot $t$, $\tau_f(t)$ represents the generation timestamp of the packet being transmitted over link at slot $t$. 
For the $s_f$, the AoI is reset upon packet generation. For any intermediate node or $d_f$, the AoI evolution depends on the packet reception from its incoming links. 
%

Let $x_{i,j,f}(t) \in \{0,1\}$ be the binary decision variable, indicating if link $l_{i,j}$ transmits a packet of flow $f$ at slot $t$.
Following the rigorous formulation in \cite{Wang2019AOI}, the AoI evolution at node $j$ is governed by the activity of its incoming link $l_{i,j}$, $A_{j,f}(t)$ can be calculated as follows:
\begin{equation}\label{eq:aoi_eva1}
\begin{aligned}
& A_{j,f}(t)=\left[x_{i,j,f}(t-1)\left(t-\tau_{f}(t-1)\right)\right]+\left(A_{j,f}(t-1)+1\right) \\
& \times\left(1-x_{i,j,f}(t-1)\right),\forall i,j \in \mathcal{N}, \forall f \in \mathcal{F}, 1 \leq t \leq T.
\end{aligned}
\end{equation}

In multi-hop IAB network, AoI of one node is related to the AoI of the previous hop. Therefore, the AoI evolution between hops is given as follows:
\begin{equation}\label{eq:aoi_eva2}
\begin{aligned}
& A_{j,f}(t)=\left[x_{i,j,f}(t-1)\left(A_{i,f}(t-1) + 1\right)\right]+\left(A_{j,f}(t-1)+1\right) \\
& \times\left(1-x_{i,j,f}(t-1)\right),\forall i,j \in \mathcal{N}, \forall f \in \mathcal{F}, 1 \leq t \leq T,
\end{aligned}
\end{equation}

Due to the PD strategy, $d_f$ will receive the packets transmitted by two paths, the AoI is updated if either path delivers a packet. The effective AoI at the $d_f$ is:
\begin{equation}\label{eq:aoi_df}
\begin{aligned}
  A_{d_f,f}(t) = & \min \{ A_{d_f,f}(t-1)+1, \\
   & \min_{k \in \{1,2\}} ( A_{v_k,f}(t-1) + 1 \mid x_{v_k,d_f,f}(t-1)=1 ) \},
\end{aligned}
\end{equation}
where $l_{v_k, d_f}$ is the last hop link of path $p_{f,k}$, $v_k \in \mathcal{N}$.

To measure the AoI of each flow $f$, we employ the average instantaneous AoI at the $d_f$ over $T$ time slots. Therefore, the average AoI of flow $f$ can be calculated as follows:
\begin{equation}\label{eq:aoi_avg}
\Lambda(f)=\lim_{T \rightarrow \infty} \frac{1}{T} \sum_{t=1}^T \mathbb{E} [ A_{d_f,f}(t) ], \forall f \in \mathcal{F},
\end{equation}

\subsubsection{Queuing modeling}
We model the packet transmission process as a discrete-time queuing system. Let $Q_{n}^f(t)$ denote the queue length of flow $f$ at node $n$ at the beginning of time slot $t$. The queue evolution depends on the difference between the aggregate arriving packets and the scheduled departing packets, expressed as:
\begin{equation}\label{eq:queue_evol1}
\begin{aligned}
  Q_{n}^f(t+1) = & \max \{Q_{n}^f(t) + \sum_{i \in \mathcal{N}} x_{i,n,f}(t) - \sum_{j \in \mathcal{N}} x_{n,j,f}(t),0\}\\
  & , \forall n \in \mathcal{N}, \forall f \in \mathcal{F}, 1 \leq t \leq T.
\end{aligned}
\end{equation}

The adoption of the PD strategy leads to traffic flow replication and aggregation at intermediate nodes, making queue management critical. In practical deployments, network nodes are constrained by finite buffer size. Consequently, the accumulated queue backlog at any node $n$ must not exceed its physical buffer capacity threshold, denoted as $\Gamma_n$. To prevent buffer overflows and packet drops, we impose the following hard buffer constraint:
\begin{equation}\label{eq:queue_evol2}
\begin{aligned}
  \lim_{T \rightarrow \infty} \frac{1}{T} \sum_{t=1}^T \mathbb{E} [ \sum_{f \in \mathcal{F}} Q_{n}^f(t) ] \leq \Gamma_n, \forall n \in \mathcal{V} \setminus \{s_f, d_f\}, 1 \leq t \leq T.
\end{aligned}
\end{equation}

\subsubsection{Routing Constraints}
Let $y_{i,j,f}^k \in \{0,1\}$ be the binary decision variable, indicating whether the link $l_{i,j}$ is on the route path $p_{f,k}, k \in \{1,2\}$ of the transport flow $f$. The routing variable $y_{i,j,f}^k$ must satisfy the law of conservation of flow, which is expressed as:
\begin{equation}\label{eq:flow_cons}
\begin{aligned}
\sum_{i \in \mathcal{N}} y_{i,n,f}^k - \sum_{j \in \mathcal{N}} y_{n,j,f}^k & = 
\begin{cases} 
1, & n = s_f \\ 
-1, & n = d_f \\ 
0, & \text{otherwise,}
\end{cases} \\
 & \forall i,j,n \in \mathcal{N}, f \in \mathcal{F}, k \in \{1,2\}.
\end{aligned}
\end{equation}

In addition, we implement the PD strategy by enforcing node-disjoint path to promote transmission reliability, expressed as:
\begin{equation}\label{eq:node_disjoint}
\sum_{k=1}^{2} \sum_{i \in \mathcal{N}} y_{i,n,f}^k \leq 1, \forall n \in \mathcal{N} \setminus \{s_f, d_f\}. 
\end{equation}

\subsubsection{Scheduling Constraints}

Only the links on the selected route path that have packets to be transmitted will be scheduled. Therefore, the link scheduling variables $x_{i,j,f}(t)$ and routing variables $y_{i,j,f}^k$ need to be bound, expressed as:
\begin{equation}\label{eq:route_binding}
\begin{aligned}
  x_{i,j,f}(t) &\leq y_{i,j,f}^{k} \cdot \mathbb{I}(Q_{i}^f(t) > 0), \\
  & \forall i,j \in \mathcal{N}, f \in \mathcal{F}, k \in \{1,2\}, 1 \leq t \leq T, 
\end{aligned}
\end{equation}
where $\mathbb{I}(\cdot)$ is the indicator function.

Furthermore, we assume that the scheduler obtains real-time blockage status through periodic beam training and rapid link-layer feedback. Leveraging the high directionality and severe path loss of mmWave communications, we assume interference is negligible for nodes separated by a distance greater than $R$. While dense deployments may exhibit complex interference, this model effectively captures the dominant conflict-set dynamics, as mmWave networks typically operate in a noise-limited regime due to blockage and narrow beamforming \cite{Al2016mmWave}. 
Within this range, we enforce an interference avoidance scheduling constraints. By integrating this with the half-duplex constraint of IAB node, where simultaneous transmission and reception are prohibited. We construct a conflict set $\mathcal{I}_{i,j}$ for each link $l_{i,j}$. This set encompasses all potential links that either are located within the interference range $R$ or share a common node with link $l_{i,j}$. To guarantee conflict-free transmission, at most one link within the conflict set and the link $l_{i,j}$ itself can be activated simultaneously at any time slot $t$, which is constrained by:
\begin{equation}\label{eq:conflict_avoidance}
\begin{aligned}
\sum_{f \in \mathcal{F}} x_{i,j,f}(t) &+ \sum_{l_{u,v} \in \mathcal{I}_{i,j}} \sum_{f' \in \mathcal{F}} x_{u,v,f'}(t) \leq 1 \\
 & , \forall i,j,u,v \in \mathcal{N}, 1 \leq t \leq T.
\end{aligned}
\end{equation}

\subsection{Problem Formulation}
\label{sec:formulation}
To ensure timely status updates for the system, we aim to minimize the cumulative average AoI across all flows in the IAB network. Denoting $\Lambda(f)$ as the long-term average AoI of flow $f$, the optimization problem can be formulated as:

\begin{equation}\label{prob:reform}
\begin{aligned}
&(\textbf{P1}) \quad \min \sum_{f \in F}\Lambda(f)\\
&\text {s.t.} 
\qquad \text {hard buffer constraint: \quad Eq.(\ref{eq:queue_evol2}); }\\
&\qquad\quad\; \text {routing constraints: \quad Eq.(\ref{eq:flow_cons})(\ref{eq:node_disjoint}); }\\
&\qquad\quad\; \text {scheduling constraints: \quad Eq.(\ref{eq:route_binding})(\ref{eq:conflict_avoidance}). }
\end{aligned}
\end{equation}

$\textbf{P1}$ is a Mixed-Integer Nonlinear Programming (MINLP) problem due to the bilinear coupling in Eq.\eqref{eq:aoi_avg} and Eq.\eqref{eq:queue_evol1}. While linearization via Reformulated Linearization Technique (RLT) is theoretically possible, it causes a variable space explosion, rendering exact solutions computationally prohibitive. Furthermore, the problem remains NP-hard with exponential complexity. 

\subsection{Problem Reformulation}
Instead of pursuing a computationally expensive global optimal solution, we adopt the Lyapunov optimization framework to reformulate the long-term stochastic $\textbf{P1}$ into a tractable, per-slot deterministic optimization problem \cite{neely2013stochastic}. We define the quadratic Lyapunov function $L(t) \triangleq \frac{1}{2}\sum_{n \in \mathcal{V}} Q_n(t)^2$ to characterize queue stability, the drift term $\Delta(t)$ captures the tendency of queue growth, where $\Delta(t) \triangleq \mathbb{E}[L(t+1) - L(t)]$. The original problem is then transformed into minimizing the upper bound of the drift-plus-penalty term $\gamma \cdot \Delta(t) + \sum_{f \in \mathcal{F}}A_{n,f}(t), n \in \mathcal{V}$ in each time slot, where $\gamma > 0$ is a control parameter that balances the trade-off between network stability and information freshness.

Minimizing the drift term effectively drives traffic from congested nodes to downstream neighbors, while minimizing the penalty term optimizes information freshness. 
Through algebraic manipulation, the original stochastic optimization problem is decoupled into a series of deterministic sub-problems. 
At each time slot $t$, the optimal scheduling policy is obtained by solving the following Max-Weight Scheduling \textbf{P2}, subject to the instantaneous network availability and physical constraints:


This reformulated $\textbf{P2}$ is a classic Maximum Weight Independent Set (MWIS) problem, which serves as the theoretical foundation for our proposed low-complexity solution.

\begin{algorithm}[t]
\caption{Proposed Resilient and Freshness-Aware Scheduling Algorithm}
\label{alg:rfas}
\begin{algorithmic}[1]
\REQUIRE Flows $\mathcal{F}$, Conflict Graph $\mathcal{I}_{i,j}$, Buffer Capacity $\Gamma_n$.
\ENSURE Scheduling decisions $\mathbf{X}(t)$.
\FOR{each time Slot $t$}
\STATE Initialize candidate list $\mathcal{K} \leftarrow \emptyset$.
    \FOR{each link flow $f$ and $l_{i,j} \in \mathcal{P}_{f}$}
        \IF{$Q_j^f(t) \geq \Gamma_j$} 
            \STATE $W_{i,j,f}(t) \leftarrow -\infty$ \quad
        \ELSE
            \STATE $W_{i,j,f}(t) \leftarrow A_{d_f,f}(t) + \gamma \cdot (Q_i^f(t) - Q_j^f(t))$
            \STATE \textbf{if} $W_{i,j,f}(t) > 0$ \textbf{then} Add tuple $(l_{i,j}, f)$ to $\mathcal{K}$.
        \ENDIF
    \ENDFOR    
    \STATE Sort $\mathcal{K}$ in descending order of $W_{i,j,f}(t)$.
    \STATE Initialize the avoidance link set $\mathcal{\mathcal{C}} \leftarrow \emptyset$.
    \FORALL{candidate $(l_{i,j}, f)$ in $\mathcal{K}$}
        \IF{$l_{i,j}$ no conflict with $\mathcal{\mathcal{C}}$}
            \STATE $x_{i,j,f}(t) \leftarrow 1$
            \STATE $\mathcal{\mathcal{C}} \leftarrow \mathcal{\mathcal{C}} \cup \{l_{i,j}, \mathcal{I}_{l_{i,j}}\}$
        \ENDIF
    \ENDFOR
    \STATE Update $Q_{n}(t+1), \forall n \in \mathcal{V}$ and $A_{df,f}(t+1), \forall f \in \mathcal{F}$.
\ENDFOR
\end{algorithmic}
\end{algorithm}

\section{Proposed Low-Complexity Solution}

To make it applicable to the PD strategy in multi-hop IAB networks and further decrease the difficulty of solving \textbf{P2}, we implemented a decoupling strategy that decouples routing from scheduling. 
In the offline phase, two node-disjoint paths are pre-configured based on the $k$-shortest path algorithm for each flow to strictly satisfy the routing constraints $\text{Eq.(\ref{eq:flow_cons})}$ and $\text{Eq.(\ref{eq:node_disjoint})}$, ensuring that the probability of simultaneous blockage on both paths remains minimal. 
Subsequently, the online phase focuses on dynamic link scheduling. The process of the proposed RFAS algorithm is shown in Algorithm \ref{alg:rfas}. 
The goal of the algorithm is to find an AoI-driven scheduling scheme, while considering the balance of traffic and maintaining the resilience of the system. The idea of the algorithm is as follows.

For each time slot $t$, the algorithm initializes an empty candidate activation list $\mathcal{K}$. 
The scheduler evaluates the dynamic priority for every link $l_{i,j}$ associated with flow $f$ in the route path set $P_f$. 

Crucially, links targeting fully buffered nodes are assigned a weight of $-\infty$. This forced mechanism overrides the standard Lyapunov drift, ensuring strictly feasibility of the buffer constraint Eq.(\ref{eq:queue_evol2}). Only links with positive weights are then added to $\mathcal{K}$.

\begin{figure}[t!]
    \centering
    \includegraphics[width=0.92\columnwidth]{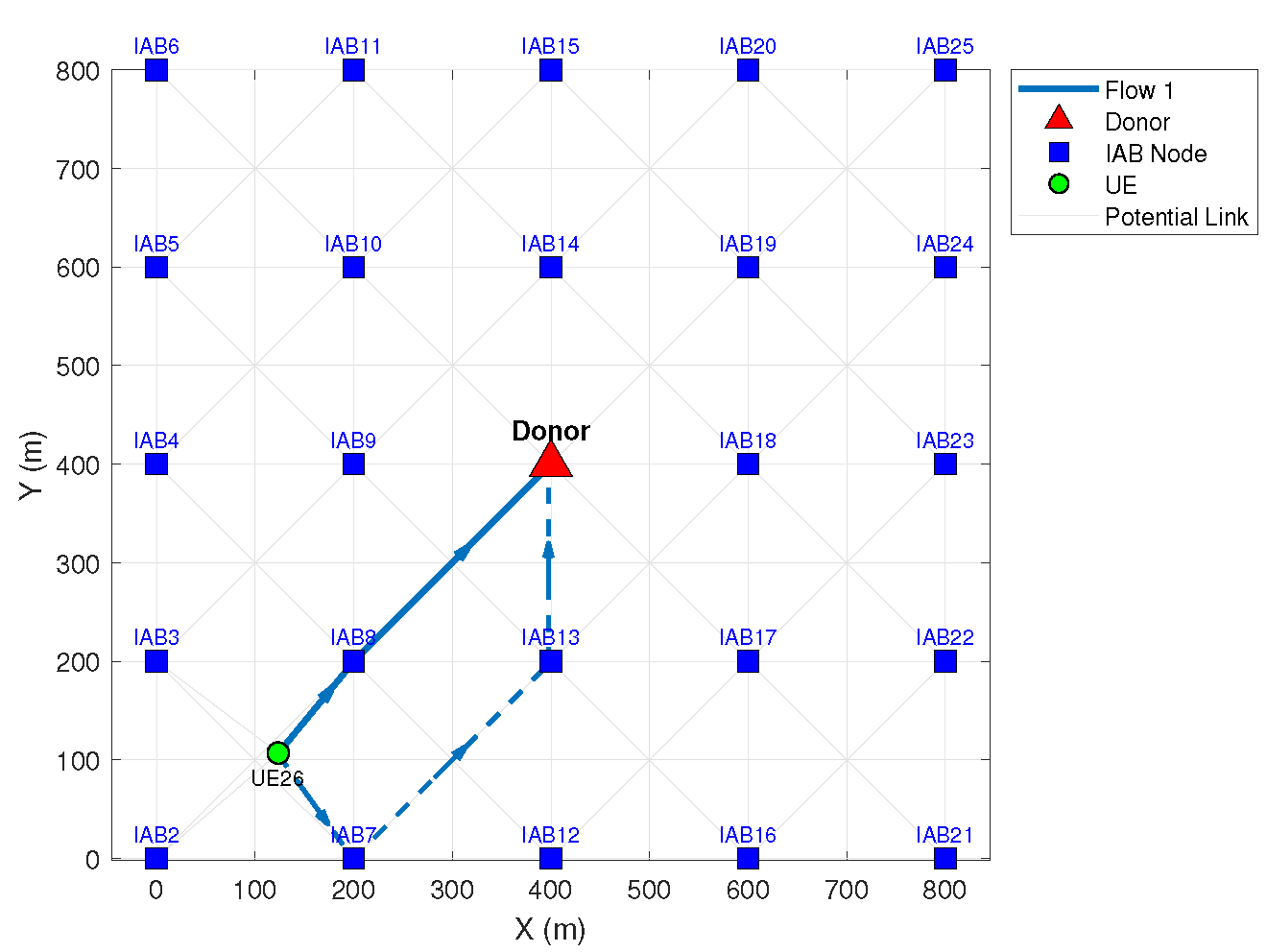} 
    \caption{Simulation topology of the industrial IAB network.}
    \label{fig:sim_topo}
\end{figure}

Next, $\mathcal{K}$ is sorted in descending order. The algorithm employs a strategy with an avoidance set $\mathcal{C}$, it iteratively activates the highest-weight link that does not overlap with $\mathcal{C}$, subsequently adding the activated link and its conflict set to $\mathcal{C}$.
Once the conflict-free schedule is finalized, the system updates the queue states and AoI accordingly. Then the algorithm will proceed to the next time slot.

Although we adopted an offline configuration of shortest paths to reduce complexity, in each time slot of the online phase, the algorithm can dynamically select the optimal available path based on the link status. In extreme cases, if all candidate paths are blocked, the queue stability constraint can ensure that data packets are buffered until a feasible path is restored, thereby maintaining the resilience of the system. 

We now show that the proposed RFAS algorithm has a polynomial time complexity, making it feasible for industrial deployment. The computational complexity of RFAS per time slot is dominated by the link sorting and conflict inspection processes. Let $L = |\mathcal{L}|$ denote the total number of links and $D_{max}$ is the maximum degree of the conflict graph. The sorting step requires $\mathcal{O}(L \log L)$, followed by the conflict-free greedy selection which scales with $\mathcal{O}(L \cdot D_{max})$. Consequently, the overall complexity is $\mathcal{O}(L \log L + L \cdot D_{max})$. This polynomial complexity represents a significant reduction compared to the exponential complexity $\mathcal{O}(2^L)$ of global Integer Linear Programming solvers, ensuring the feasibility of RFAS for millisecond industrial control loops. 

\section{Simulation Results}
\subsection{Simulation Setup}

We have constructed an industrial mmWave IAB network comprising a central IAB donor and 24 IAB nodes. These nodes are deployed in a regular grid pattern over an 800 m $\times$ 800 m factory area with an inter-node spacing of 200 meters.
The interference range $R$ is defined as 100 m. We evaluated the sensitivity of the control parameter $\gamma$, which reflects the trade-off between freshness and stability. Specifically, a larger $\gamma$ will prioritize the AoI of data packets, but it will increase the occupancy rate of the buffer;  while a smaller $\gamma$ can improve the stability of the queue, but it will lead to higher AoI; therefore, through experiments, we determined that $\gamma = 0.5$ is the optimal ideal balance point, which can minimize AoI while avoiding buffer overflow or system instability.

To capture the dynamic nature of industrial obstacles, we model the channel using a Gilbert-Elliot Discrete-Time Markov Chain \cite{3GPPTR38901R18}. Each link transitions between Line-of-Sight and Blocked states, where the steady-state blockage probability is denoted as $P_{blk}$, with a default value of 15\%. Considering the physical transit time of dynamic obstacles, the average blockage duration is set to 100 slots. 
Traffic generation follows a periodic pattern characteristic of industrial control applications. The High-mode implies a packet-generation cycle of 10 slots, the Low-mode implies a cycle of 50 slots, and in the Mixed-mode, half of the packets have a cycle of 10 slots and the other half have a cycle of 50 slots. 

The performance of the proposed RFAS algorithm is compared against three baselines. Firstly, there was a single-path scheme without considering PD. Additionally, there were two algorithms based on the MAX-weight scheduling concept, including the Queue-Aware Scheduling (QAS) scheme \cite{Hai2018DBP}, which prioritizes queue stability and load balancing , and the Freshness-Aware Scheduling (FAS) scheme \cite{Wang2019AOI}, which prioritizes AoI. 
By comparing RFAS with these baselines, we can effectively capture the trade-off between reliability and freshness scheduling, thereby conducting a comprehensive assessment of the performance of the proposed algorithm. 

\subsection{Performance Analysis}

\subsubsection{Resilience Analysis}

\begin{figure}[t!]
    \centering
    \includegraphics[width=0.98\columnwidth]{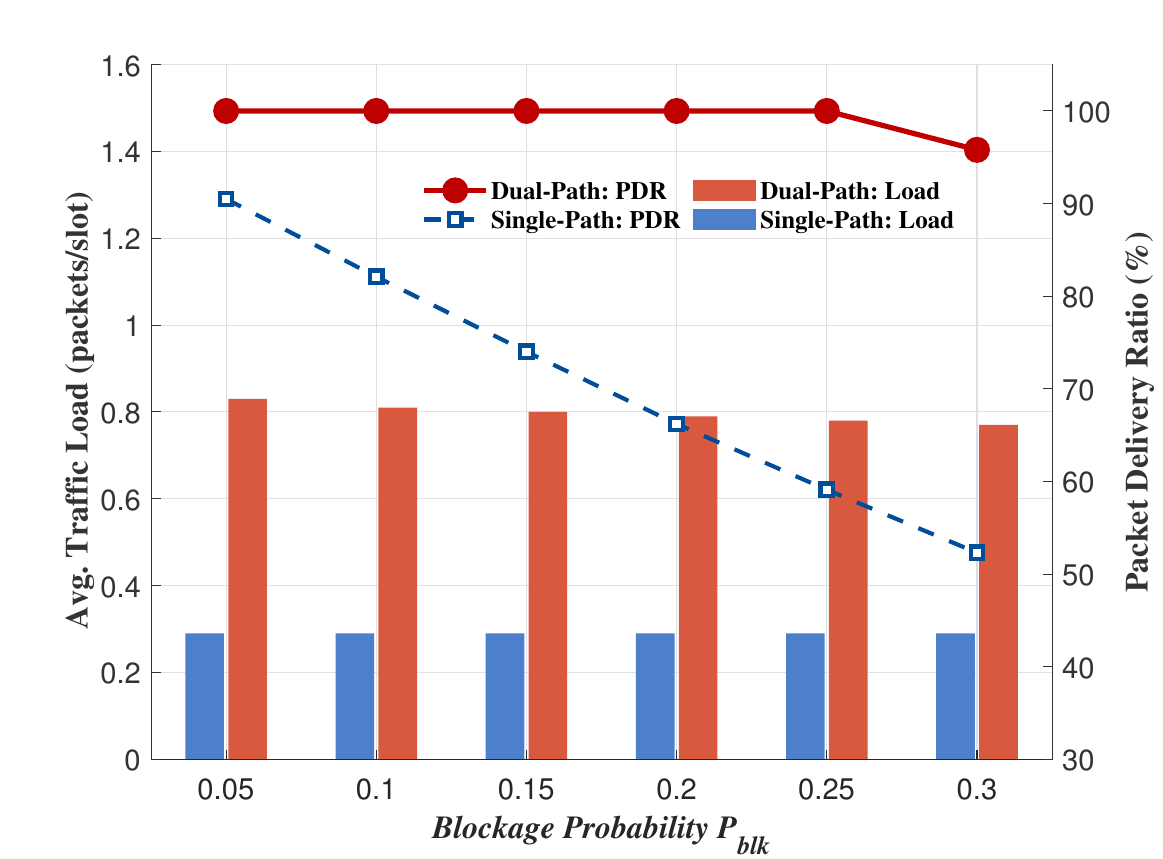}     
    \caption{Reliability analysis: PDR performance under varying blockage probabilities $P_{blk}$.}
    \label{fig:resilience}
\end{figure}
To verify the impact of PD and dual-path transmission on network resilience, we evaluate the RFAS algorithm under both Single-Path and Dual-Path routing strategies, as illustrated in Fig. \ref{fig:resilience}. The steady-state blockage probability $P_{blk}$ varies from $5\%$ to $30\%$, results indicate that the Single-Path implementation suffers significant performance degradation in harsh channel environments, losing nearly half of the status updates at high blockage rates despite utilizing an optimal scheduler. In contrast, the Dual-Path implementation maintains a robust packet delivery ratio exceeding 95\% by exploiting link diversity, effectively enhancing the network resilience. Regarding network overhead, the Dual-Path approach naturally incurs a higher traffic load due to the strategy of PD compared to the Single-Path solution. Therefore, when applying the PD strategy in industrial multi-hop IAB networks, RFAS must handle the resulting congestion while maintaining network resilience to ensure timely information transmission. This performance will be further demonstrated below. 

\subsubsection{Micro-Dynamics and Stability Trade-off}

%

We introduce a strict buffer capacity constraint of $\Gamma_n = 8$ packets, representing finite buffer size in low-cost IAB nodes. We analyze the transient response to a sudden traffic burst injected at $t=20$. 
Since FAS only focuses on reducing the negative impacts of AoI, it is unable to promptly clear the backlog queue, as shown in . Therefore, although FAS has a theoretically novel advantage, it is not applicable to resource-constrained industrial multi-hop IAB network scenarios.
In contrast, the proposed RFAS can effectively enforce strict constraints. Once the queue length approaches the upper limit, the system triggers forced transmission to alleviate congestion. This mechanism successfully limits the worst-case queue length, achieving a load balancing comparable to the QAS. 
Crucially,  indicates that this strict stability enforcement incurs a lower cost. RFAS can ensure physical stability while reducing AoI.

\subsubsection{Scalability and Industrial Robustness}

%

We verified the scalability of the algorithm and its performance in handling actual industrial traffic models. 
The results of scalability.
When the number of UEs increased from 4 to 14, the AoI of QAS significantly surged by approx $300\%$, while FAS maintained the lowest AoI, but it had a serious load imbalance problem, and the standard deviation of the queue length peaked at 3.5.
In contrast, RFAS maintained a balanced performance. It effectively suppressed queue congestion while yielding an AoI within $10\%$ of the FAS lower bound. It is important to note that FAS achieved its theoretical minimum AoI only by disregarding the hard buffer constraints, which inevitably leads to physical overflows. From the standard deviation of approximately 2.8, RFAS improves load balancing by $25\%$ compared to FAS, it can be seen that compared to methods that solely focus on freshness, RFAS has significant improvements in load balancing. 

\begin{figure}[t]
    \centering
    \subfloat[Industrial: AoI]{
        \includegraphics[width=0.48\linewidth]{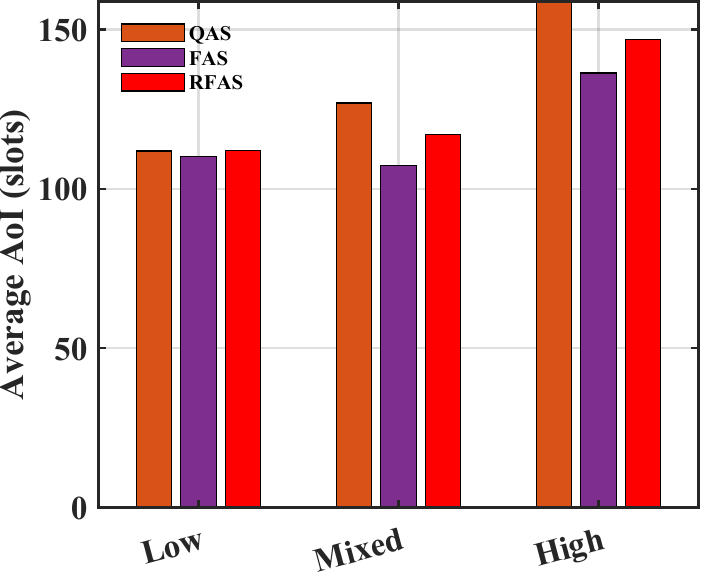}
        \label{fig:flow_a}
    }
    \hfil
    \subfloat[Industrial: Imbalance]{
        \includegraphics[width=0.45\linewidth]{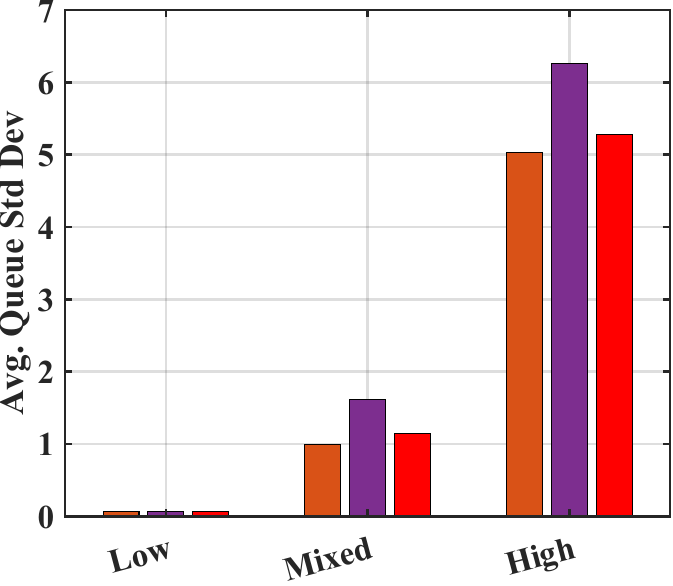}
        \label{fig:flow_b}
    }
    
    \caption{Industrial Performance evaluation showing QAS, FAS, and RFAS. (a) AoI; (b) Imbalance.}
    \label{fig:industrial}
\end{figure}

In actual industrial scenarios, there may be multiple business cycles. Through testing of different mixed cycle traffic scenarios, the superiority of RFAS has been further demonstrated. As shown in Fig.\ref{fig:flow_a} and Fig.\ref{fig:flow_b}, QAS has the highest AoI due to short-sighted scheduling decisions. However, RFAS minimizes the AoI while evenly distributes the traffic load across available resources. Therefore, RFAS exhibits similar freshness characteristics to FAS, but outperforms FAS in terms of queue stability by up to $19\%$. The balance of RFAS highlights its applicability in industrial environments, as maintaining stability and reducing AoI are equally important in these environments.

\balance
\section{Conclusion}
In this paper, a comprehensive scheduling framework has been proposed for sustainable and resilient industrial IAB networks. By decoupling routing and scheduling and adopting a composite metric derived from Lyapunov optimization, we formulated a MILP problem for minimizing AoI, and proposed a low-complexity RFAS algorithm. 
Simulation results show that RFAS effectively reconcile the contradiction between reliability and congestion caused by PD, guaranteeing queue stability under hard constraints while minimizing the average AoI compared to benchmark algorithms.
In the future, we will further expand on this research to obtain the theoretically optimal solution in the constrained mmWave environment, thereby establishing a theoretical foundation for large-scale slot optimization.



\end{document}